\documentstyle[prd,preprint,aps]{revtex}
\begin{document}
\tighten
\draft
\preprint{}

\title{Scattering of weakly interacting massive particles from ${}^{73}$Ge}
\author{ V. Dimitrov}
\address{Faculty of Physics, Sofia University, 1126 Sofia, Bulgaria
\cite{Vesko} \\
and \\
Bartol Research Institute, University of Delaware, Newark, Delaware 19716}
\author{J. Engel}
\address{Deptartment of Physics and Astronomy, CB3255,
University of North Carolina,
Chapel Hill, North Carolina  27599}
\author{S. Pittel}
\address{Bartol Research Institute, University of Delaware, Newark, Delaware
19716 }
\date{\today}
\maketitle
\begin{abstract}
We use a ``hybrid" method, mixing variationally-determined triaxial nuclear
Slater determinants, to calculate the response of ${}^{73}$Ge to
hypothetical dark-matter particles such as neutralinos. The method is a hybrid
in that rotational invariance is approximately restored prior to
variation and then fully restored before the mixing of intrinsic states.
We discuss such features of
${}^{73}$Ge as shape coexistence and triaxiality, and their effects on
spin-dependent neutralino cross sections.
Our calculations yield a satisfactory
quadrupole moment and an accurate magnetic moment in this very complicated
nucleus, suggesting that the spin structure and thus the axial--vector
response to dark matter particles is modeled well.
\end{abstract}

\vspace{0.25in}


The identity of the invisible stuff believed to constitute most of the material
universe is still unknown\cite{r:DM1,r:DM2}.  For several years now, heavy
weakly-interacting particles (WIMPs) have been an attractive
candidate\cite{r:DM2}.  Although some recent evidence indicates the possible
presence of macroscopic objects in the galactic halo\cite{r:machos}, statistics
are poor and conclusions uncertain, and the WIMP hypothesis is still quite
plausible.  A variety of experiments to detect WIMPs are in fact either already
operating or in the planning/prototype stage.  Among the most promising is a
germanium-based detector that incorporates the odd-mass isotope
${}^{73}$Ge\cite{r:ge73}, which carries spin.  While scalar (spin-independent)
cross sections for ``neutralinos", perhaps the most plausiby motivated WIMP,
now
appear usually to be larger than the axial-vector (spin-dependent) cross
sections in this nucleus\cite{r:Drees}, there are still regions of parameter
space for which this is not so.  Furthermore, WIMPs with no scalar
interactions,
such as heavy Majorana neutrinos (perhaps with reduced coupling to the $Z$),
have not yet been completely ruled out.  A careful investigation of the
spin-dependent response of ${}^{73}$Ge is therefore desirable.
Spin-independent
scattering can be easily calculated following, e.g., the work of Ref.\
\cite{r:Engel}

Several papers have addressed aspects of spin-dependent scattering from
${}^{73}$Ge.
Engel and Vogel\cite{r:EV} used data from magnetic moments to
estimate the quenching of the neutron spin in several heavy nuclei, including
germanium.  Iachello, Krauss and Maino\cite{r:IBFM} employed the Interacting
Boson Fermion Model, and Nikolaev and Klapdor-Kleingrothaus\cite{r:Klapdor}
used
finite-fermi-systems theory to calculate the same quantities (Ref.\
\cite{r:IBFM} also included a calculation of the small proton spin).  The most
comprehensive study of ${}^{73}$Ge to date appears in Ref.\ \cite{r:Ressell}
where a large--basis shell--model calculation was performed and the full
spin-dependent neutralino response, including the finite-q form
factors\cite{r:Engel}, was calculated.  Here we present an alternative, equally
comprehensive calculation that, we argue, is better than any of the above and
which leads to significantly different results from those obtained so far.

The principal difficulty in describing the spin
of ${}^{73}$Ge is its complicated collective structure.  A few
years ago, in a series of papers\cite{r:VAMPIR1}, the EXCITED VAMPIR
method\cite{r:VAMPIR} was used to explore the even--isotopes of germanium.
${}^{72}$Ge turned out to be particularly interesting and complicated.  The
authors concluded that both oblate and prolate shapes were represented in the
ground state wave function, a scenario that most likely persists in
${}^{73}$Ge.

As difficult as it is to describe collective shape coexistence in even--even
nuclei, it is even more so in odd--mass systems.  Shell--model methods have a
hard time incorporating enough mixing of spherical configurations to properly
describe such dynamics, particularly in odd--mass or odd--odd systems.  While
the IBFM can incorporate the dominant collective effects, it has trouble
including the spin polarization that plays a crucial role in
axial-vector scattering.  In addition, it cannot be readily applied at non-zero
momentum transfer.  The VAMPIR method, though in principle well suited for
such problems, cannot at present be used in odd--mass systems.

The method we use here is described in detail in Ref.\ \cite{r:Dimitrov}.  It
shares with VAMPIR the idea of mixing variationally determined
Slater determinants, in which symmetries are broken but restored either before
or after variation.  In our calculation the symmetries broken in the intrinsic
states are those associated with rotational invariance, parity, and axial
shape.  Ideally we would like to restore all these symmetries before
variation; unfortunately that is too expensive computationally at present.
Our ``hybrid" procedure is to restore axial symmetry,
parity invariance, and approximate rotational invariance
(using a method similar to that proposed by Kamlah\cite{r:Kamlah}) prior to
the variation of each intrinsic state, and then subsequently to fully restore
rotational invariance before mixing the intrinsic states.

The procedure differs from VAMPIR, in that it allows fully triaxial Slater
determinants at the expense of particle--number
breaking\cite{r:DS}.  Our recent work\cite{r:Dimitrov}, along with that of
other groups using completely different methods\cite{r:Draayer},
indicates that the trading of number nonconservation for triaxiality is a
good idea, despite the apparent loss of pairing correlations
traditionally associated with the former.  Pairing forces
evidently induce effective triaxiality.  Though the precise relationship of
triaxiality to pairing needs to be further clarified, numerical
results\cite{r:Dimitrov} in, e.g., the 0s,1d shell show that our approach is
as accurate and efficient as VAMPIR for describing even--even systems while
also providing a reliable reproduction of the collective dynamics of odd--mass
systems, something VAMPIR cannot yet do precisely because of its BCS--like
treatment of pairing.

In the calculations for ${}^{73}$Ge that we report below, we assume a
single--particle space for both protons and neutrons consisting of the full
$0f,1p$ shell plus the $0g_{9/2}$ and $0g_{7/2}$ orbitals.  Our goal is
to include all of the single--particle orbits that
are important for low--energy properties of the nucleus ${}^{73}$Ge plus
all of their spin--orbit partners. Despite its
size --- no space this large can be fully treated in the shell model ---
our space imposes only a modest burden on the computer
programs that implement the hybrid method.

The size of our space does, however, lead to one feature that is not
present in the 0s,1d--shell tests reported in Ref.\ \cite{r:Dimitrov}.
There all single--particle levels have the same parity.  In our
current work, the single--particle basis includes levels of both positive and
negative parities, allowing parity invariance to be broken.  At first
glance, this may seem like an unfortunate complication; in fact it is a
benefit.  Consider a system of
four identical particles interacting via a pairing force acting in two closely
related model spaces.  The first contains two degenerate single--particle
levels, the $f_{7/2}$ and the $f_{5/2}$, with the same parities; the second
contains the levels $g_{7/2}$ and $f_{5/2}$, with the same angular momenta as
in the first but now with opposite parities.  The exact ground--state energies
in these two models are identical.  A mixing of triaxial mean--field states
can always describe the ground state of the two systems, the only issue being
how many intrinsic states are required.  We have carried out calculations for
both models with the clear conclusion that convergence is more
rapid in the second.  The reason is also clear:  the freedom to break (and
subsequently restore) another symmetry permits us to build more correlations
into the intrinsic states.

One caveat accompanies parity mixing:  ideally it should be done
democratically.
Put another way, all of the dominant single--particle orbits should be allowed
to benefit from parity mixing.  At the practical level, this means that two
complete oscillator shells should be included in a parity--mixed calculation.
Since we are not currently able to include so large a single--particle space
for
germanium, we will somehow have to simulate the missing effects in our
analysis.
We will discuss how we do this shortly.

Returning to germanium, we note that a crucial ingredient in any realistic
nuclear--structure calculation is an
appropriate nuclear hamiltonian.  The one-- and two--body parts must be
compatible with one another and also with the model space.  This is difficult
to
achieve; microscopic two--body interactions, derived for example from a
G-matrix, include monopole pieces that are unable to describe the movement of
spherical single--particle levels as one passes from the beginning to the end
of
a shell\cite{r:mono,r:iodine}.  In several papers, including a very recent one
on neutrino scattering from iodine\cite{r:iodine}, we proposed a procedure for
avoiding this problem.  It consists basically of removing from the two--body
interaction {\it all} monopole components and shifting their effects to the
single--particle energies.  We follow the same procedure here; our two--body
force, for example, is a fit to a Paris--potential G-matrix\cite{r:paris},
modified as just described.

To determine spherical single--particle energies, we first carry out BCS fits
with the above force to the spherical quasi--particle energies in the mass-71
isotopes ${}^{71}$Ga and ${}^{71}$Zn.  These are the odd--mass nuclei closest
to
${}^{73}$Ge that are nearly spherical.  For levels very far from the Fermi
surface, where quasiparticle energies are ambiguous, we estimate the
single--particle energies from general systematics.  Since our model space does
not include two full oscillator shells, however, we have found it necessary to
adjust slightly some of the BCS--produced single--particle energies.  The
$0f_{7/2}$ and $0f_{5/2}$ orbits strongly mix with the $0g_{9/2}$ and
$0g_{7/2}$
orbits, respectively,
which have opposite parity,  but our space does not include
positive--parity orbits to mix with the $1p_{3/2}$ and $1p_{1/2}$.  We
therefore
adjust the single--particle energies of the $1p$ levels to simulate the missing
levels; our criterion is a reasonable reproduction of the occupation numbers
obtained in the VAMPIR calculations of ${}^{72}$Ge.  The end result is the set
of single--particle energies shown in Table I.  We should note here that these
energies are different from and more reasonable than those used in Ref.\
\cite{r:Ressell}.  We believe the difference can be attributed to insufficient
correlations in the shell model calculation \cite{r:Ressell},
and to our removal of monopole forces.

The results of our calculations involve only the $J^{\pi}=9/2^+$ ground state;
for technical reasons\cite{r:Dimitrov}, higher-lying states are not as well
modeled.  We have included in our analysis four intrinsic states, whose
properties are summarized in Table II.  The absolute energies are meaningless
but the energy differences between configurations are important.  The variables
$\beta$ and $\gamma$ are the usual radial and angular measures of deformation
in
the collective model\cite{r:BM} (given here for both protons and neutrons).
They imply
that the lowest state is predominatly oblate and slightly triaxial, and the
second predominantly prolate and also slightly triaxial.
These two intrinsic states mix very little with one
another, however.  This is a bit surprising since there seems to be significant
oblate--prolate mixing in the VAMPIR results for ${}^{72}$Ge.  The other two
intrinsic states, both predominantly oblate, do mix somewhat more strongly into
the ground state, lowering its energy by 0.95 MeV.  A summary of important
quantities in the final mixed ground state is given in Table III.

The most important result, the ground--state magnetic dipole moment, is in
excellent agreement with the experimental value.  Such good agreement is
difficult to achieve.  The Schmidt magnetic moment (arising from a pure
$0g_{9/2}$ neutron configuration) is -1.91 {\em n.m.}, very far from the
experimental value of -0.88 {\em n.m.}.  Ressell and
collaborators\cite{r:Ressell}, in their ``large--space" shell--model
calculation, were able to quench the magnetic moment significantly
to -1.24 {\em n.m.},
but could not account for the remaining difference.  Our calculation,
despite the small number of intrinsic states, contains the full quenching
required by experiment, a point on which we elaborate shortly. (It is
interesting to note that we already achieve a
magnetic moment of $\mu=-0.909$ {\it n.m.}  with just the lowest intrinsic
state.)

Our calculated quadrupole moment is not quite as good (the experimental
value is
-21 $e-fm^2$), though still reasonable.  The discrepancy is probably an
indication that there should be somewhat more mixing of the prolate solution in
our ground state, which would take us the right direction.  Of course it is
also
possible that the inclusion of additional intrinsic states could modify the
quadrupole moment.

To clarify the differences between our magnetic moment and that of Ref.\
\cite{r:Ressell}, we compare in Table IV the results of both approaches for the
various spin and orbital angular momenta that contribute.  The most important
difference occurs in the neutron spin, for which our result is significantly
smaller.  The large and negative neutron spin g--factor ($g^s_n=-3.826$) makes
this the chief source of our improved result.  The differences in the spins,
unlike those in the orbital angular momenta, carry over into WIMP scattering
cross sections.

Magnetic moments can be modified in two fundamentally different ways ---
through
nuclear structure correlations not included in the model calculation, and
through meson exchange corrections and/or nucleonic resonances.  The
non--nuclear contributions, though certainly important in a detailed theory of
magnetic moments, rarely affect the final results by more than about 10\%.
Since our goal is a reliable description of the nuclear spin structure, we aim
at (and achieve) agreement with experiment moment to within 10\%.  In the work
of Ref.\ \cite{r:Ressell}, as noted above, the magnetic moment is too large by
roughly 30\%.  It appears to us that most of this discrepancy is due to the
omission of important nuclear stucture correlations.  To compensate, Ref.\
\cite{r:Ressell} advocates quenching the isovector spin only (note:  this is
not
done correctly everywhere in the paper; the two nominally equivalent
prescriptions for quenching neutralino cross sections are in fact different),
using the observation that the isovector spin is more strongly quenched from
{\it its Schmidt value} than the isoscalar spin.  It is not obvious, however,
that the same statement should be true when the spins have already been partly
quenched by nuclear correlations.  Our results indicate that quenching both
the isoscalar and isovector ``large--space" spins would have been the
best procedure in
Ref.\ \cite{r:Ressell}.  Of course, such a prescription should not be blindly
extended to non-zero momentum transfer\cite{r:DM3}.  Our calculations, since
they apparently correctly represent the spin structure, require no quenching at
$q=0$ and no arbitrary assumptions about how the form factor should change at
$q
\neq 0$.  For these reasons we believe our neutralino cross sections, to which
we now turn, to be the most reliable yet obtained.

In Fig.~1, we present the three functions that determine the spin-dependent
cross sections for any neutralino at all momementum transfers, written in terms
of $y=( b q/2)^2$, where $b=A^{1/6}~\rm{fm} = 2.04$ fm is
the oscillator parameter
(the precise definitions of the three functions are given in refs.\
\cite{r:Engel,r:DM3}).  Comparing the results for $S_{00}(y)$ (the pure
isoscalar form factor) and $S_{11}(y)$ (the isovector form factor) with the
corresponding large--space results of Ref.\ \cite{r:Ressell} (see Fig.~4 of
that
reference), we conclude that {\it both} are reduced relative to theirs.
Quenching the isovector cross sections alone does not seem appropriate at any
momentum transfer.

To allow use of these functions (or form factors) we have made polynomial fits
to them.
They are very well represented over the full range of
relevant momenta by the following sixth--order polynomials:
\begin{eqnarray*}
S_{00}(y) &=& ~0.1606 ~-~1.1052~y ~+~ 3.2320~y^2
{}~-~4.9245~y^3 \\
&+&~4.1229~y^4 ~-~1.8016~y^5 ~+~0.3211~y^6 \\
S_{11}(y) &=& ~0.1164 ~-~0.9228~y ~+~ 2.9753~y^2
{}~-~4.8709~y^3 \\
&+&~4.3099~y^4 ~-~1.9661~y^5 ~+~0.3624~y^6 \\
S_{01}(y) &=& ~-~0.2736 ~+~2.0374~y ~-~ 6.2803~y^2
{}~+~9.9426~y^3 \\
&-&~8.5710~y^4 ~+~3.8310~y^5 ~-~0.6948~y^6 \\
\end{eqnarray*}

In Fig.~2, we show the full axial form factor for ``B-ino" scattering from
${}^{73}$Ge, the same choice considered in Ref.\ \cite{r:Ressell}.  We assume
that the spin structure of the nucleon is as given by the original EMC
experiment, which, we should note, has recently been called into
question\cite{r:Kamio}.  For
comparison, we also show the Independent Single Particle Shell Model (ISPSM)
result, derived under the assumption that all the spin is carried by
one $0g_{9/2}$ neutron.  Importantly,
nuclear correlations significantly quench
the ISPSM form factor at {\it all} momentum transfers.  The quenching at $q=0$
by a factor of almost three persists out to very large momentum transfers and
in
fact even grows slightly stronger.  This most likely reflects the prominent
role
played by the $0g$ orbits in building the correlations that lead to spin
quenching.

Several improvements to our analysis should be considered in the future.  The
most important is an explicit incorporation of the remaining orbits from the
2s,1d,0g shell, whose effects were treated very roughly here.  This
should enable us to remove some of the arbitrariness in the single--particle
energies that resulted from our incomplete treatment of parity--mixing effects.
It may also lead to a slight lowering of the prolate intrinsic state, thereby
permitting more mixing with the dominant oblate state.  Shell-model
calculations
have more or less reached their limit for the time being; the improvements just
outlined above, by contrast, can be incorporated with reasonable amounts of
analysis, coding, and computer time.  Should the spin-response of ${}^{73}$Ge
be
needed more accurately, we will implement the improvements.

We wish to acknowledge useful conversations with M.T.\ Ressell.  This work was
supported in part by the U.S.\ Department of Energy under grants
DE-FG05-94ER40827, by the National Science Foundation
under grants PHY-9108011, PHY-9303041 and INT-9224875, and by the Bulgarian
Scientific Foundation under Contract $\Phi$210/2090.

\begin{figure}
\caption{The calculated functions $S_{00}$, $S_{01}$ and $S_{11}$
for ${}^{73}$Ge, as a function of $y=(bq/2)^2$ ($b$ is the
harmonic-oscillator length parameter).
The solid line is $S_{00}$, the dashed line is $S_{11}$ and the
dotted line is $S_{01}$. These functions are
defined in Refs. [3,21].}
\label{f:fig1}
\end{figure}

\begin{figure}
\caption{The calculated spin structure function $S(y)$ for pure
$\tilde{B}$ scattering from ${}^{73}$Ge, assuming EMC couplings.
The solid curve gives the results of the calculation
described in the text. The dashed curve gives the ISPSM results.}
\label{f:fig2}
\end{figure}

\begin{table}
\caption{Single--particle energies (in MeV)
used in the calculation of the structure
of ${}^{73}$Ge described in the text.
\label{t:spes}}
\begin{tabular}{ccccc}
& orbit & protons & neutrons  \\
\tableline
&  $0f_{7/2}$ &      0   &    0   &\\
&  $1p_{3/2}$ &      1.5 &    4.2 &\\
&  $0f_{5/2}$ &      4.0 &    4.0 &\\
&  $1p_{1/2}$ &      3.3 &    5.3 &\\
&  $0g_{9/2}$ &      5.1 &    6.4 &\\
&  $0g_{7/2}$ &     13.0 &   13.5 &\\
\end{tabular}
\end{table}

\vspace{1in}

\begin{table}
\caption{Properties of the four intrinsic states
for ${}^{73}$Ge that are used in the calculations
described in the text. The shape parameters
are obtained in the intrinsic frame; the energy and
magnetic moment in the lab frame.
\label{t:intrinsic}}
\begin{tabular}{cc|cc|cccc}
& Intrinsic~State & Energy (MeV) & $\mu$ & $\beta_p$
& $\gamma_p$ (deg.) &$\beta_n$ &$\gamma_n$ (deg.) \\
\tableline
& 1 &    48.27  & -0.909 & .14  & 50.3 & .08 & 55.3   \\
& 2 &    49.38  & -0.279 & .10  & 14.2 & .05 & ~1.7  \\
& 3 &    49.71  & -0.816 & .12 & 53.2 & .07 & 50.3 \\
& 4 &    49.88  & -0.857 & .12 & 54.5 & .08 & 55.3 \\
\end{tabular}
\end{table}

\vspace{1in}

\begin{table}
\caption{Nuclear structure properties of the ground state
of ${}^{73}$Ge.
\label{t:evalues}}
\begin{tabular}{cccc}
& Energy~(MeV)&    ~47.32 &\\
& $\mu$~({\em n.m.}) &  ~~-0.920 &\\
& Q~ ($e-fm^2$)&   -40.98  &\\
\end{tabular}
\end{table}

\vspace{1in}

\begin{table}
\caption{Comparison of spin and angular momenta of ${}^{73}$Ge that
derive from our ``hybrid" variational approach and the large--basis
shell--model approach of Ref. [10].
\label{t:angmom}}
\begin{tabular}{cccccc}
&& $<S_{p}>$ & $<S_{n}>$ & $<L_{p}>$ & $<L_{n}>$ \\
\tableline
 & Present~Work&   .030    &  .378 & .361 & 3.732 \\
 & Ref.\ \cite{r:Ressell} &   .011   &  .491 & .468 & 3.529  \\
\end{tabular}
\end{table}

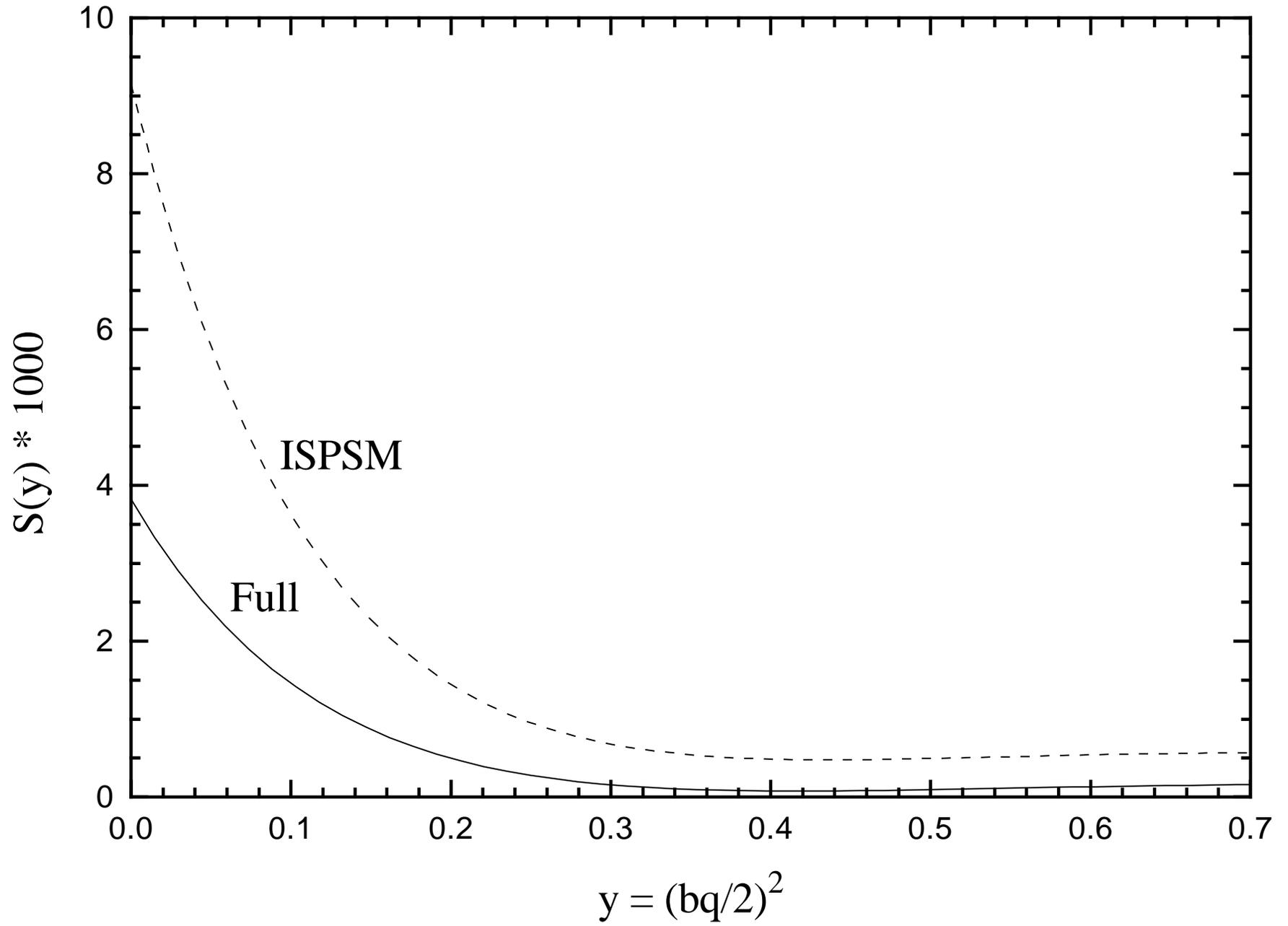
\end{document}